\documentclass[12pt]{iopart}
\usepackage{iopams}
\usepackage{graphicx}

\begin{document}

\title[Thermal expansion and magnetostriction of RAgSb$_2$]
{Thermal expansion and magnetostriction of pure and doped RAgSb$_2$ (R = Y, Sm, La) single crystals.}

\author{S L Bud'ko$^1$, S A Law$^1$\footnote{currently at Department of Physics, University of Illinois at Urbana-Champaign}, P C Canfield$^1$, G D Samolyuk$^1$, M S Torikachvili$^2$ and G M Schmiedeshoff$^3$}

\address{$^1$Ames Laboratory US DOE and Department of Physics and Astronomy,
Iowa State University, Ames, IA 50011, USA}

\address{$^2$Department of Physics, San Diego State University, San Diego, CA 92182, USA}

\address{$^3$Department of Physics, Occidental College, Los Angeles, CA 90041, USA}

\begin{abstract}
Data on temperature-dependent, anisotropic thermal expansion in pure and doped RAgSb$_2$ (R = Y, Sm, La) single crystals are presented. Using the Ehrenfest relation and heat capacity measurements, uniaxial pressure derivatives for long range magnetic ordering and charge density wave transition temperatures are evaluated and compared with the results of the direct measurements under hydrostatic pressure. In-plane and $c$-axis pressure have opposite effect on the phase transitions in these materials, with in-plane effects being significantly weaker. Quantum oscillations in magnetostriction were observed for the three pure compounds, with the possible detection of new frequencies in SmAgSb$_2$ and LaAgSb$_2$. The uniaxial (along the $c$-axis) pressure derivatives of the dominant extreme orbits ($\beta$) were evaluated for YAgSb$_2$ and LaAgSb$_2$.

\end{abstract}


\submitto{\JPCM}

\maketitle

\section{Introduction}

The RAgSb$_2$ series of compounds crystallizes in a simple tetragonal ZrCuSi$_2$-type structure ($P4/nmm$, No.
129) \cite{bry95a,sol95a}. Members of the family show rich and complex electronic and magnetic properties,
including charge density wave (CDW) transitions in LaAgSb$_2$ \cite{mye99a,son03a}, anisotropic, ferromagnetic,
Kondo-lattice behavior in CeAgSb$_2$ \cite{mye99a,hou95a,mur97a,sid03a,job05a} and low temperature, crystalline
electric field (CEF) governed metamagnetism in RAgSb$_2$ compounds with R = heavy rare earth. \cite{mye99a,mye99b}
Recent increased attention to this family is partially due to the successful growth of high quality single
crystals \cite{mye99a} that are suitable for detailed, anisotropic thermodynamic and transport measurements as
well as for studies of the Fermi surfaces (FS) of these materials through measurements of quantum oscillations.
\cite{job05a,mye99c,ina02a,pro07a}.

In this work we report measurements of anisotropic thermal expansion (TE) and longitudinal ($H \| L \| c$)
magnetostriction (MS) for pure members of the series: non-magnetic YAgSb$_2$ and LaAgSb$_2$ and antiferromagnetic
(below $\sim 8.6$ K SmAgSb$_2$) as well as two samples in which La is partially substituted with either Ce
(Ce$_{0.2}$La$_{0.8}$AgSb$_2$) or Nd (Nd$_{0.25}$La$_{0.75}$AgSb$_2$). YAgSb$_2$ behaves as a rather simple,
normal metal with no phase transitions observed at ambient pressure below the room temperature \cite{mye99a}.
Temperature-dependent resistivity and magnetic susceptibility measurements on LaAgSb$_2$ show two features, a stronger one at $T_1 \approx 210$ K and
more subtle one at $T_2 \approx 185$ K \cite{mye99a,son03a}. The features in resistivity are reminiscent of charge
density wave (CDW) transitions.  An X-ray scattering study \cite{son03a} revealed that indeed both features are
the signatures of CDW orderings, with the one at $T_1$ corresponding to a development of periodic charge/lattice
modulation along the $a$-axis with the wave vector $\tau_1 \sim 0.026(2\pi/a)$ and the one at $T_2$ marking an
additional CDW ordering along the $c$-axis with the wave vector $\tau_2 \sim 0.16(2\pi/c)$. Both CDW orderings
were shown to be consistent with the enhanced nesting in the different parts of the LaAgSb$_2$ Fermi surface \cite{son03a}.
The higher temperature CDW transition was shown to be very sensitive to pressure and/or rare-earth-site substitution
\cite{bud06a,tor07a,wat06a}. In both doped samples the higher temperature charge density wave transition is suppressed down to
$\sim 110$ K, additionally, in Ce$_{0.2}$La$_{0.8}$AgSb$_2$ single-ion-Kondo-like behavior in the resistivity is
observed at low temperatures and a ferromagnetic transition is detected at $\approx 3.2$ K \cite{bud06a,tor07a}. It
should be mentioned that the pressure derivatives of the higher temperature CDW transition temperatures in the two
doped samples differ by more than a factor of two from each other \cite{tor07a}.

Since the anisotropic thermal expansion data in the RAgSb$_2$ family are available only for CeAgSb$_2$
\cite{adr03a,tak03a} (data for polycrystalline LaAgSb$_2$ are also presented in Ref. \cite{adr03a}) we deem it to be
desirable, specifically for detailed analysis of the results similar to one in Ref. \cite{tak03a} to have
experimental TE data for a non-magnetic analogue. Furthermore, anisotropic pressure derivatives of CDW, Curie and
N\'{e}el transition temperatures may be estimated by combining heat capacity and anisotropic TE data,
potentially shedding some light on the reason for the variation of the $d T_{CDW}/d P$ for the different materials under
study.

\section{Experimental methods and computational details}

Plate-like RAgSb$_2$ crystals were solution grown \cite{can92a} from Sb-rich self-flux (see Refs.
\cite{mye99a,mye99b} for further details on growth procedure). Thermal expansion data were obtained using a capacitive
dilatometer constructed of OFHC copper; a detailed description of the dilatometer is presented elsewhere
\cite{sch06a}. The dilatometer was mounted in a Quantum Design PPMS-14 instrument and was operated over a
temperature range of 1.8 to 300 K in magnetic field up to 140 kOe. The samples were cut and polished so as to
have parallel surfaces perpendicular to the $[100]$ and $[001]$ directions with the distances $L$ between the surfaces
ranging between $0.5 - 2$ mm. Heat capacity of the samples was measured using a hybrid adiabatic relaxation
technique of the heat capacity option in a Quantum Design PPMS-14. Field dependent magnetization for several samples was measured using a Quantum Design MPMS-7 SQUID magnetometer. The pressure dependence of the N\'{e}el temperature
of SmAgSb$_2$ was measured by following, as a function of pressure, the sharp feature (caused by loss of
spin-disorder scattering at $T_N$) in the in-plane resistance. Pressure was generated in a Teflon cup filled with
60:40 mixture of n-pentane and light mineral oil inserted in a 33 mm outer diameter, non-magnetic,
piston-cylinder-type, Be-Cu pressure cell with a core made of NiCrAl (40 KhNYu-VI) alloy. The pressure at room
temperature was monitored by a manganin, resistive pressure gauge. At low temperatures the pressure value was
determined from the superconducting transition temperature of pure lead \cite{eil81a}. The temperature environment for
the pressure cell was provided by a PPMS instrument. Near the N\'{e}el transition the temperature was changed at
0.5 K/min rate and stabilized at every measured point (so that the effective rate was about 0.2 K/min). An additional Cernox sensor, attached to the body of the
cell, served to determine the temperature of the sample for these measurements (the temperature difference between the PPMS sensor and the sensor on the cell depends both on the temperature range and on the nominal cooling/warming rate and in our was ranging between few tenths of a degree at low temperatures and few degrees near room temperature).

The electronic structure of YAgSb$_2$ was calculated using the atomic sphere approximation, tight binding linear
muffin-tin orbital (TB-LMTO-ASA) method \cite{and75a,and84a} within the local density approximation (LDA) with
Barth-Hedin \cite{bar72a} exchange-correlation at experimental values of the lattice parameters $c/a_0 = 2.4525$ and under conditions
of uniaxial stress $c/a_+ = c/a_0 + \Delta = 2.5015$ and uniaxial pressure $c/a_- = c/a_0 - \Delta = 2.4034$ ($\Delta = 2\%$ of the $c/a$ value). The unit cell volume and the sizes of the atomic spheres were kept constant in the calculations. A self-consistency of the potential was obtained using 637 $\vec{k}$ points in the irreducible part of the Brillouin zone. The Fermi surface was calculated using 18081 $\vec{k}$ points.

\section{Results and discussion}

\subsection{YAgSb$_2$}
\begin{figure}[tbp]
\begin{center}
\includegraphics[angle=0,width=120mm]{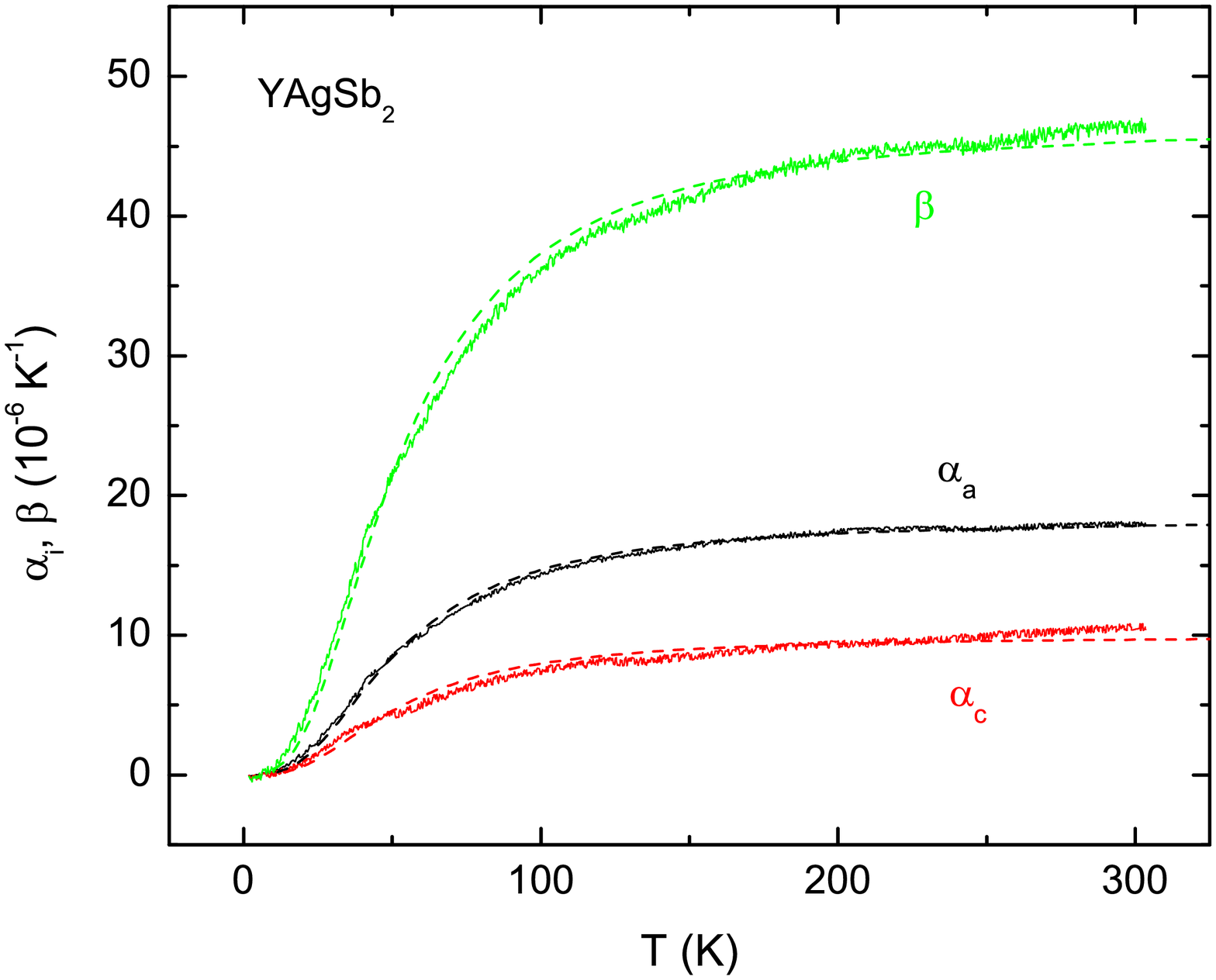}
\end{center}
\caption{(Color online) Anisotropic, temperature-dependent, linear and volume thermal expansion of YAgSb$_2$. Dashed lines show Debye fits with $\Theta_D = 215$ K.} \label{F1}
\end{figure}

The anisotropic thermal expansion of YAgSb$_2$ is shown in Fig. \ref{F1}. In-plane thermal expansion is larger than that along the $c$-axis by factor of $\sim 1.8$. The data can be represented, reasonably well (dashed lines in Fig. \ref{F1}), within a simple Debye approximation \cite{bar99a} with a temperature-independent Gr\"{u}neisen parameter (in many materials the value of the Gr\"{u}neisen parameter is approximately $1 - 2$ \cite{kri79a}). The Debye temperature, $\Theta_D = 215$ K, used in this fit is well within the range of the $\Theta_D$ values evaluated for other RAgSb$_2$ (R = rare earth) materials \cite{mur97a,ina02a,tak03a}.

\begin{figure}[tbp]
\begin{center}
\includegraphics[angle=0,width=90mm]{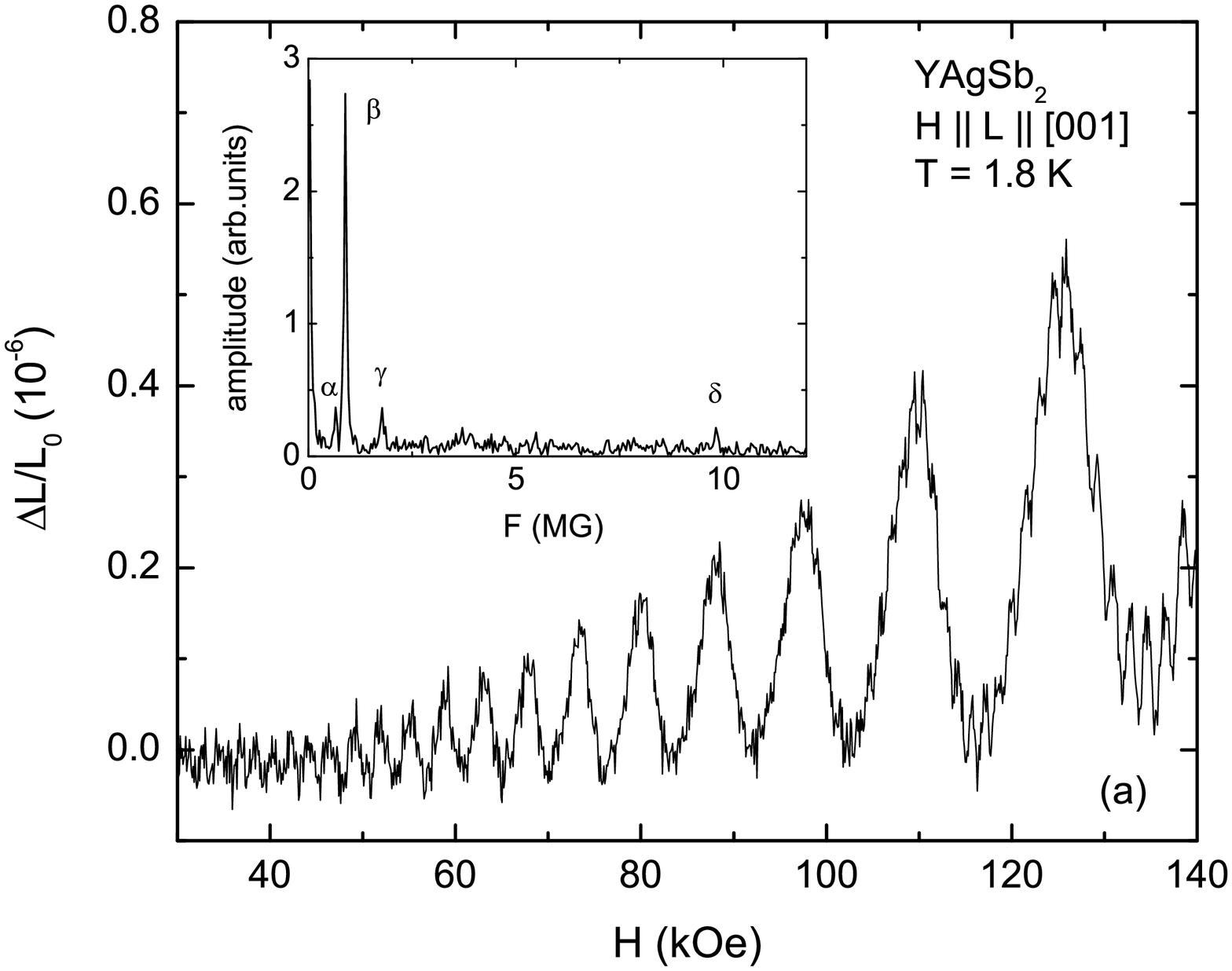}
\includegraphics[angle=0,width=90mm]{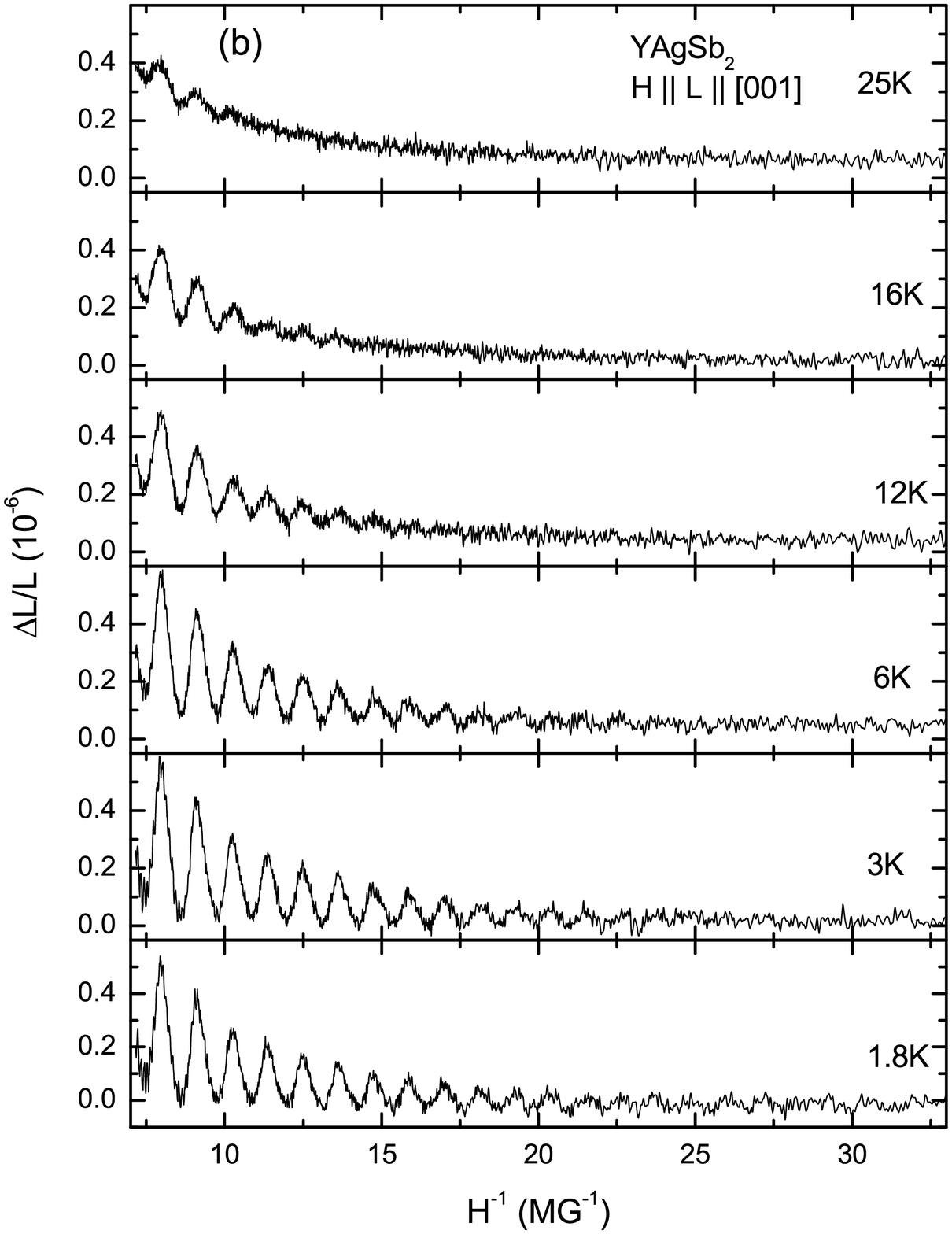}
\end{center}
\caption{Oscillations in longitudinal magnetostriction of YAgSb$_2$ ($H \| L \| [001]$). (a) $\Delta L/L_0 (H)$ taken at $T = 1.8$ K. inset - fast Fourier transform of the same data in $\Delta L/L_0 (1/H)$ form, the observed frequencies are labeled in accordance with Ref. \cite{mye99c}. (b) quantum oscillations of magnetostriction at temperatures up to 25 K plotted as a function of $1/H$.} \label{F2}
\end{figure}

Magnetostriction of YAgSb$_2$ at the base temperature is rather small, $|\Delta L/L_0 (H)| < 0.5 \cdot 10^{-6}$ at 140 kOe for both orientations of the applied magnetic field. When magnetic field is applied along the $c$-axis, clear de Haas - van Alphen (dHvA) - like oscillations of the MS are observed (Fig. \ref{F2}). These oscillations are observed up to at least 25 K. A  fast Fourier transform (FFT) of these data allows for identification of four frequencies (Fig. \ref{F2}a, inset), consistent with those observed in torque and magnetization \cite{mye99c}. The effective masses corresponding to these frequencies are consistent with those reported earlier \cite{mye99c}. The occurrence of quantum oscillations in MS is a known phenomenon \cite{cha71a}, however observations of such oscillations are rather rare,
since both large, high quality single crystals and sensitive dilatometers are required. The amplitude of the MS oscillations along the $i$-axis, $\epsilon_i$, can be written as \cite{cha71a}
\begin{displaymath}
\epsilon_i = - M H \frac{\partial \ln S_m}{\partial \sigma_i}
\end{displaymath}
where $M$ is the amplitude of the oscillations in magnetization, $H$ is a magnetic field (we will not distinguish between $H$ and $B$ for the materials studied in this work), $S_m$ is the extremal cross-sectional area of the Fermi surface perpendicular to the direction of the applied field and $\sigma_i$ is the stress along the $i$-axis. From the equation above one can see that the oscillations in MS can be used to study Fermi surfaces of metals on a par with more traditional quantum oscillations in magnetization and magnetoresistance. Due to the additional factor, $\partial \ln S_m / \partial \sigma_i$, the orbits with high sensitivity to stress have a chance to be resolved easier by magnetostriction measurements (and vice versa, stress-insensitive orbits can be easily missed). Finally, if the same orbit is detected in both, magnetization and MS measurements (preferably on the same sample), the stress derivative of the extremal cross-sectional area of the Fermi surface can be estimated. Such estimates are potentially very useful since the direct measurement of Fermi surfaces under uniaxial stress is difficult and rare.

\begin{figure}[tbp]
\begin{center}
\includegraphics[angle=0,width=120mm]{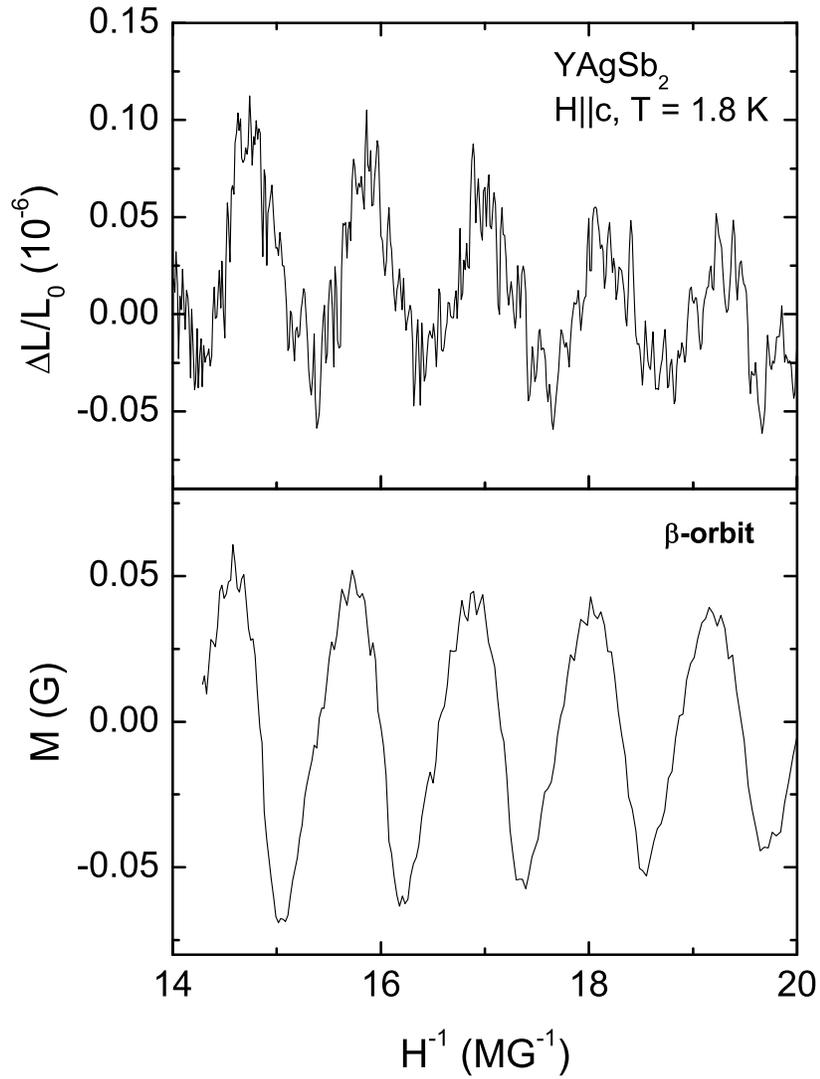}
\end{center}
\caption{Quantum oscillations in YAgSb$_2$ ($H \| [001], T = 1.8$ K as measured by magnetostriction (upper panel) and magnetization (lower panel).} \label{F3}
\end{figure}

Fig. \ref{F3} shows the oscillations corresponding to the $\beta$ orbit (in notation of Ref. \cite{mye99c}) as seen by MS and magnetization. Using the equation above, the uniaxial stress derivative for this orbit is estimated to be $\partial \ln S_{\beta} / \partial \sigma_c = - 16 \cdot 10^{-12}$ cm$^2$/dyne. In principle, one can obtain the uniaxial stress derivatives of the non-dominant frequencies by comparing the corresponding FFT amplitudes, in such case care should be taken to determine the relative phase of the magnetization and MS oscillations which defines the sign of the stress derivative.

\begin{figure}[tbp]
\begin{center}
\includegraphics[angle=270,width=90mm]{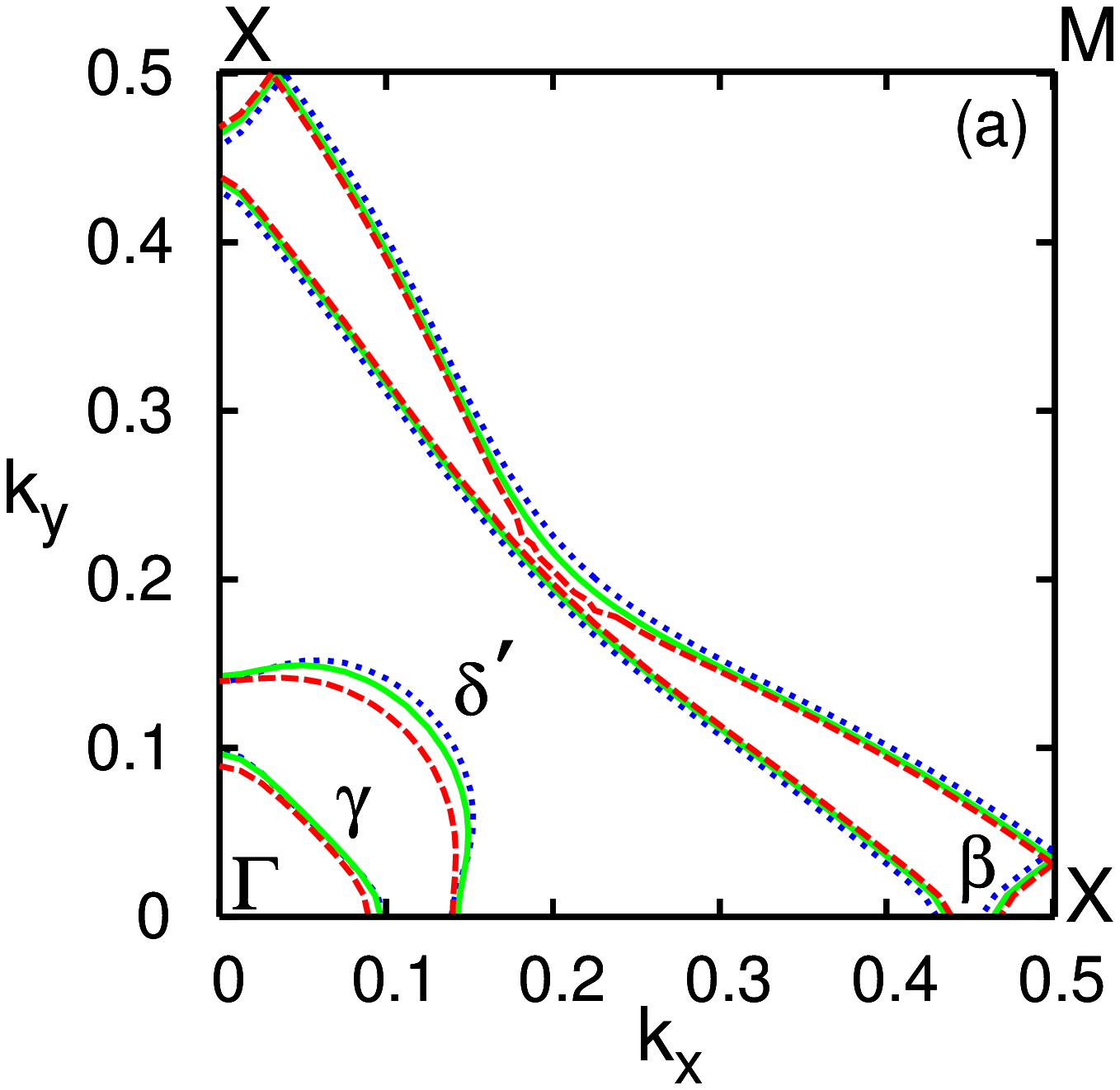}
\includegraphics[angle=270,width=90mm]{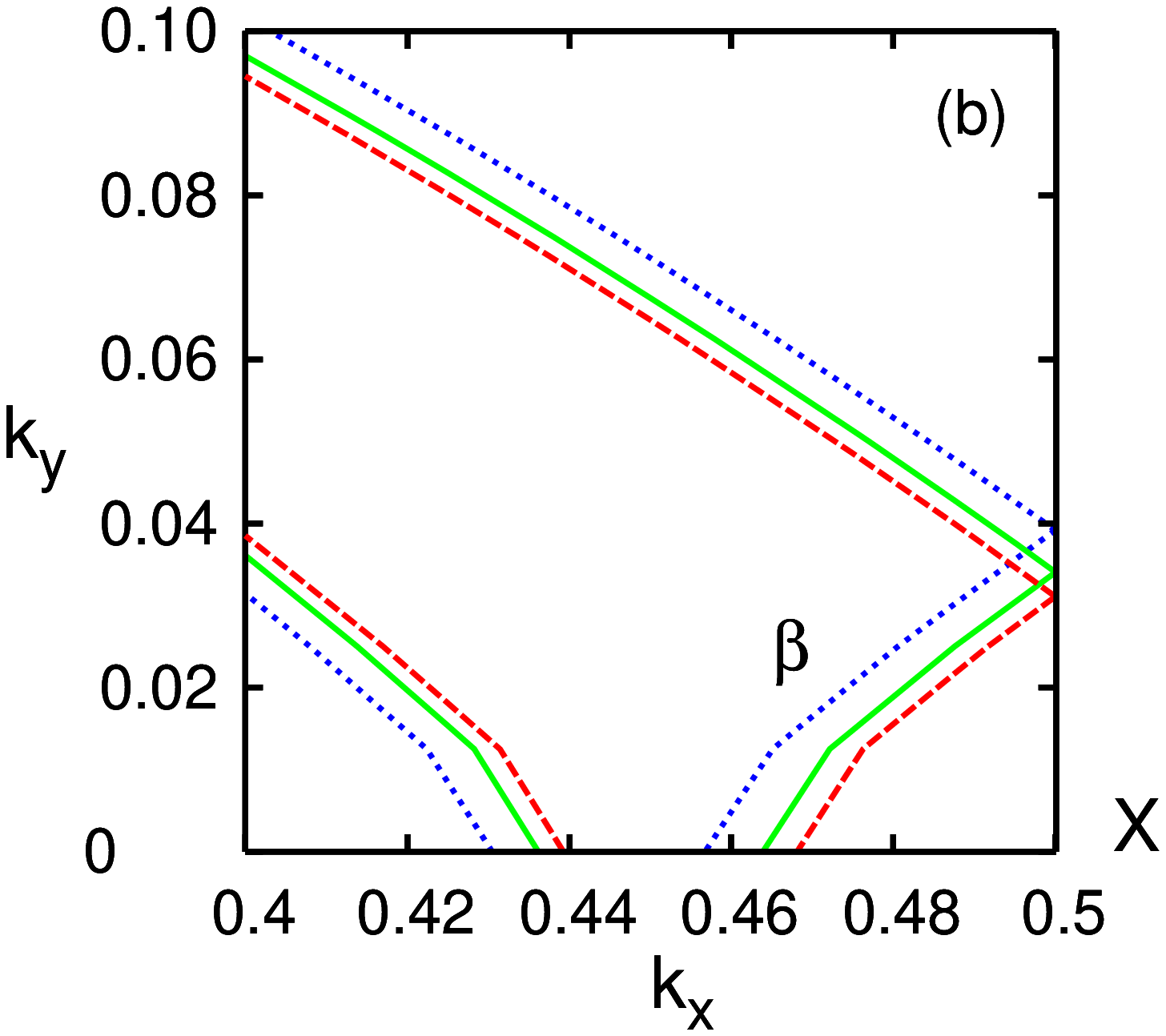}
\includegraphics[angle=270,width=90mm]{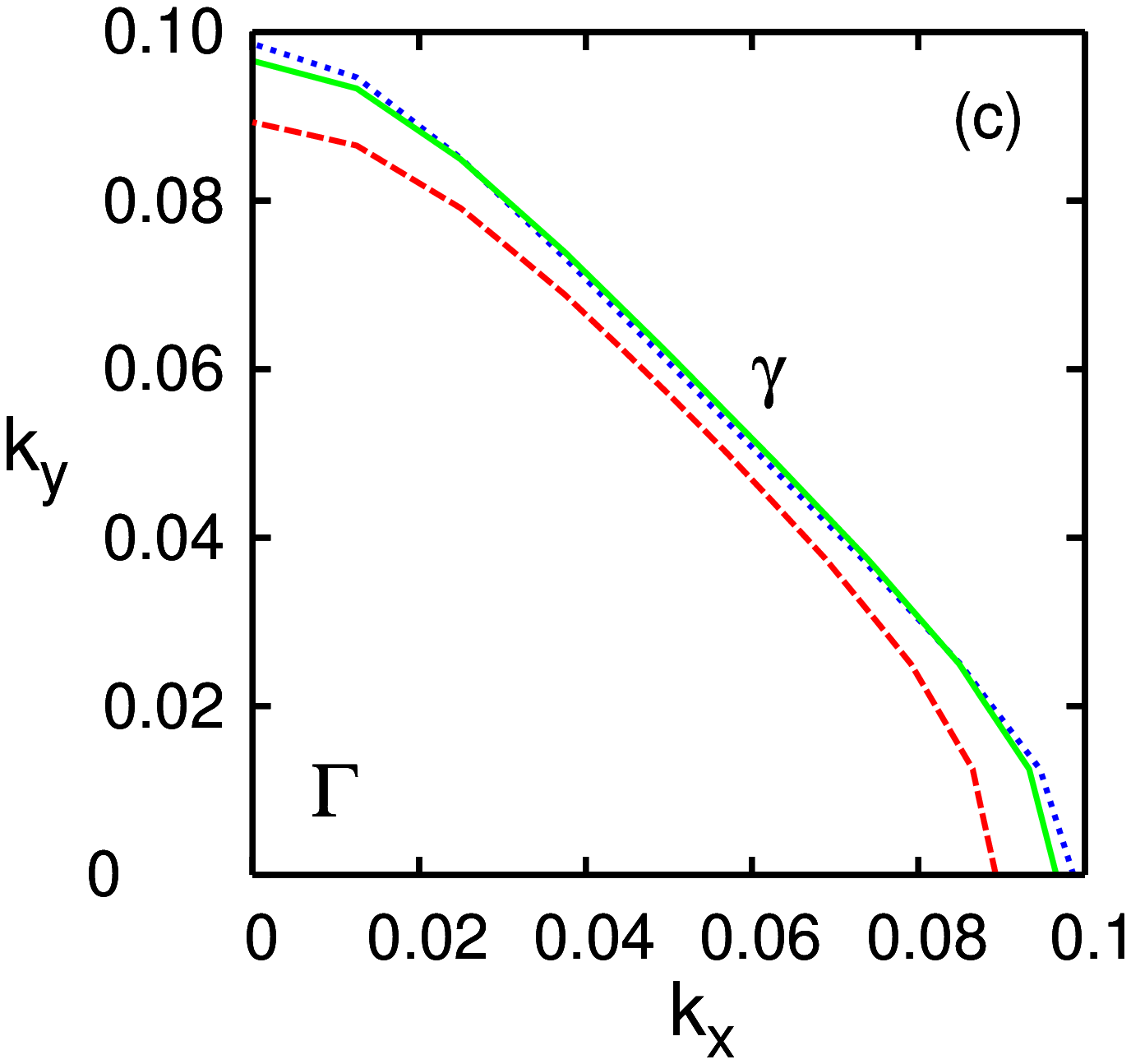}
\end{center}
\caption{(Color online) (a) $\Gamma-X-M$ cross-section of the Fermi surface of YAgSb$_2$; (b) enlarged part near $X$-point; (c) enlarged part near $\Gamma$-point. Cross-sections of the Fermi surface are labeled in agreement with the notation in Ref \cite{mye99c}. Solid (green) line corresponds to experimental $c/a_0$, dashed (red) to $c/a_+$, dotted (blue) to $c/a_-$.} \label{F4}
\end{figure}

The calculated $\Gamma-X-M$ cross-section of the YAgSb$_2$ Fermi surface for different $c/a$ values is shown in Fig. \ref{F4}. These calculations are consistent with the previous publications \cite{mye99c}. Cross-section areas of several Fermi surface sheets, $\beta$, $\gamma$ and $\delta'$, increase under uniaxial pressure along the $c$-axis ($c/a_-$) and decrease under uniaxial stress along the $c$-axis ($c/a_+$). Qualitatively, the computational results for $\beta$ orbit are consistent with the aforementioned experimental data. Quantitative comparison between the band structure calculations and the experiment requires use of the elastic constants tensor for YAgSb$_2$ that is not known at this point.

\subsection{SmAgSb$_2$}

\begin{figure}[tbp]
\begin{center}
\includegraphics[angle=0,width=100mm]{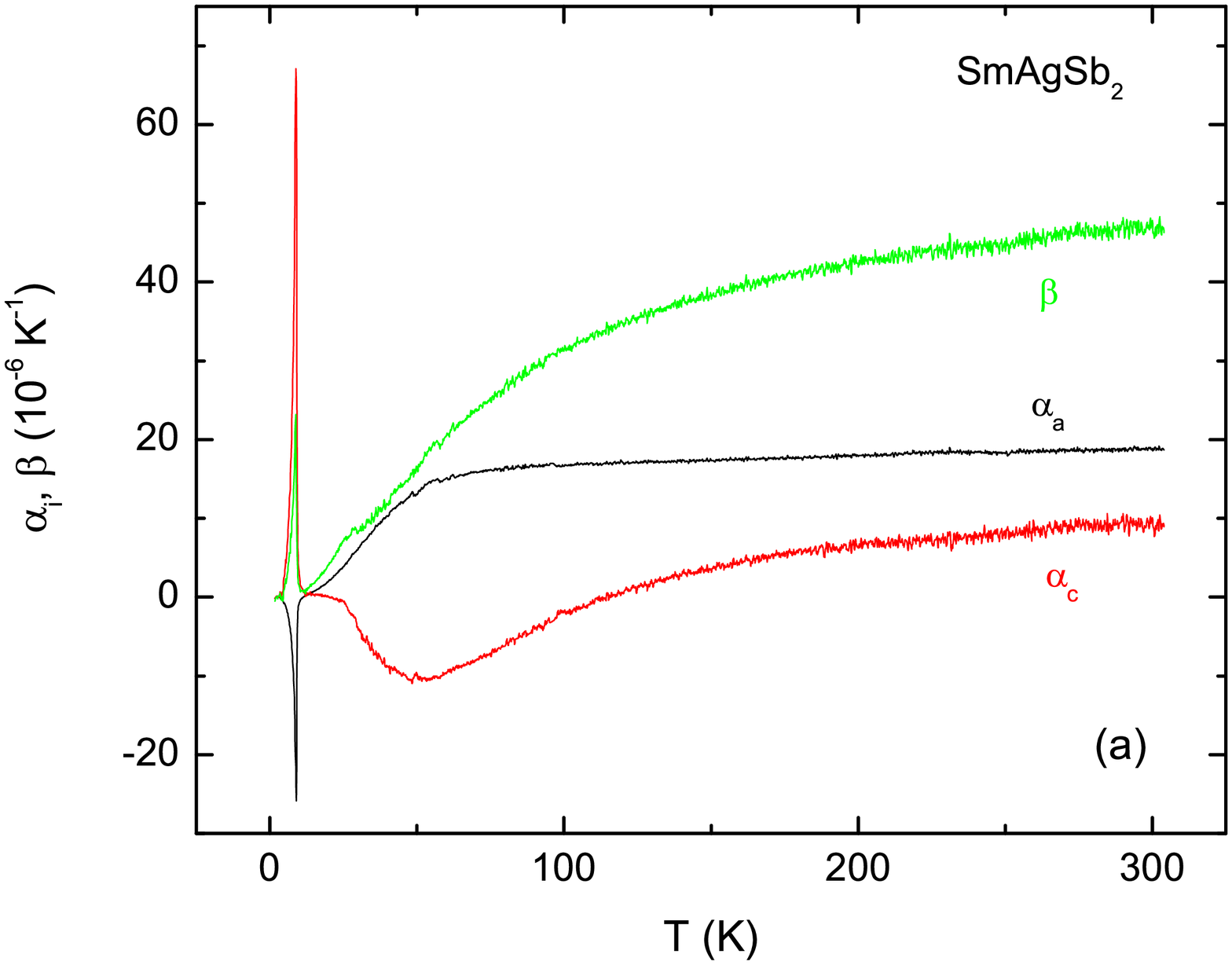}
\includegraphics[angle=0,width=100mm]{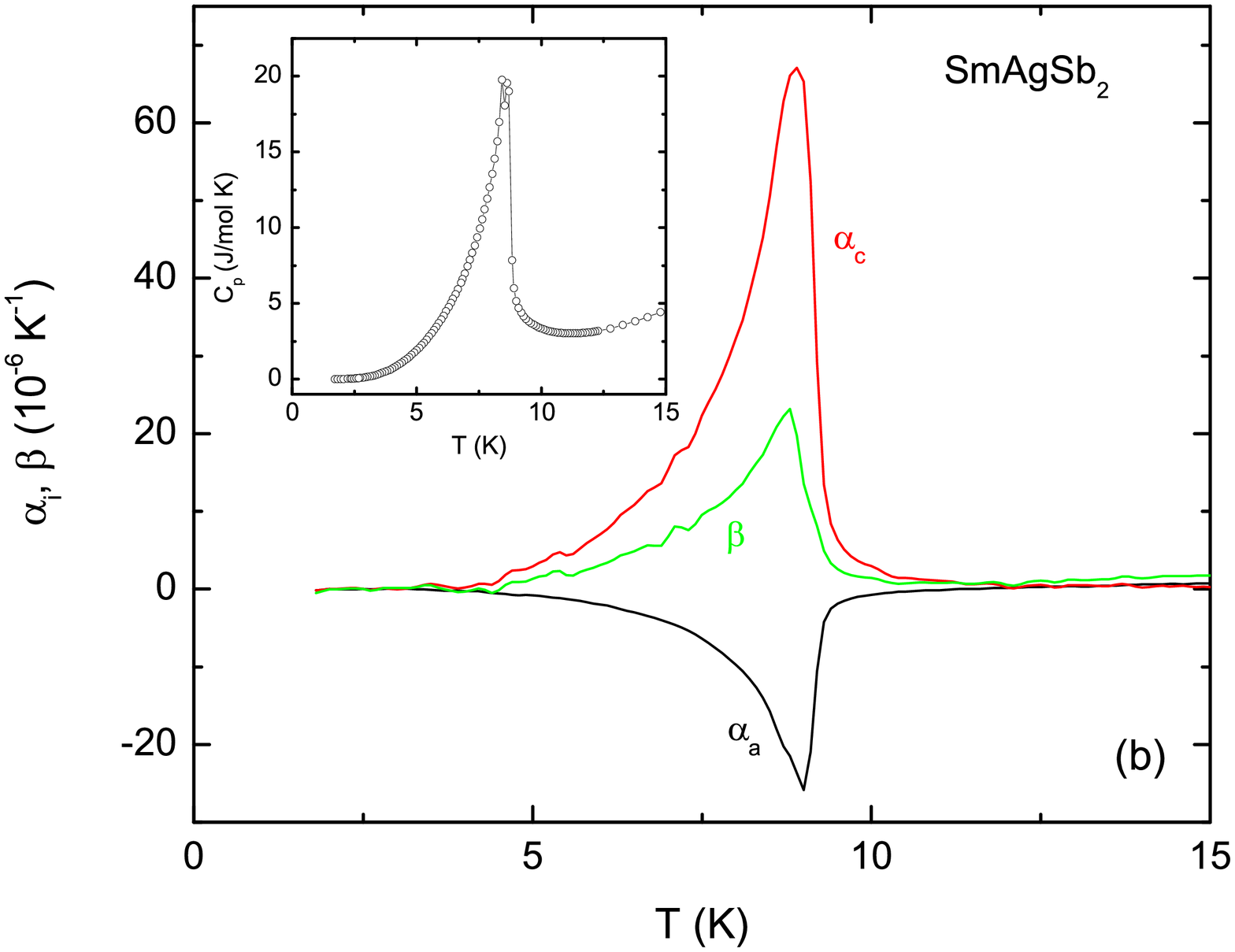}
\end{center}
\caption{(Color online) (a) Anisotropic temperature-dependent linear and volume thermal expansion of SmAgSb$_2$. (b) Enlarged low temperature part of the graph in panel (a). Inset to (b) - low temperature heat capacity data.} \label{F5}
\end{figure}

The anisotropic, temperature-dependent thermal expansion of SmAgSb$_2$ is shown in Fig. \ref{F5}. Near room temperature the thermal expansion of SmAgSb$_2$  is similar to that of YAgSb$_2$ (Fig. \ref{F1}). On cooling, $\alpha_c(T)$ changes sign to negative below $\sim 115$ K, passes through a broad minimum around 50 K and then changes sign again at $\sim 20$ K. The behavior of $\alpha_a(T)$ between the room temperature and $\sim 20$ K is less dramatic, although the difference from the $\alpha_a(T)$ behavior in YAgSb$_2$ is clearly seen below $\sim 100$ K. These features in the  SmAgSb$_2$  are possibly related to the crystalline electric field (CEF) effects (c.f. with the data of CeAgSb$_2$ \cite{tak03a}). At the same time the volume thermal expansion $\beta(T)$ in the $\sim 20 - 300$ K temperature range is similar for both materials (Figs. \ref{F1},\ref{F5}). Clear, $\lambda$-shaped, features associated with the long range order antiferromagnetic transition at $T_N \approx 8.7$ K \cite{mye99a} are seen in heat capacity and linear and volume thermal expansion (Fig. \ref{F5}(b)). The peak in $\alpha(T)$ is negative for the measurements along the $a$-axis and positive along the $c$-axis, that sums up to a (smaller) positive peak in volume thermal expansion $\beta(T)$, these signs are opposite to the ones observed at the ferromagnetic transition in CeAgSb$_2$ \cite{tak03a}. The initial uniaxial pressure derivatives of the second order phase transitions can be estimated using the Ehrenfest relation \cite{bar99a}:
\begin{displaymath}
dT_{crit}/dp_i = \frac{V_m \Delta\alpha_i T_{crit}}{\Delta C_P}
\end{displaymath}
where $V_m$ is the molar volume, $\Delta\alpha_i$ is a change of the linear ($i = a, c$) or volume ($\alpha_V = \beta$) thermal expansion coefficient at the phase transition, and $\Delta C_P$ is a change of the specific heat at the phase transition. Using experimental values: $V_m = 1.181 \cdot 10^{-4}$ m$^3$/mol, $T_N \approx 8.7$ K, $\Delta\alpha_a \approx - 2.6 \cdot 10^{-5}$ K$^{-1}$, $\Delta\alpha_c \approx 7.3 \cdot 10^{-5}$ K$^{-1}$, $\Delta\beta \approx 2.3 \cdot 10^{-5}$ K$^{-1}$, and $\Delta C_P \approx 18.7$ J/mol K, we can estimate initial uniaxial and hydrostatic pressure derivatives of the N\'eel temperature in SmAgSb$_2$: $dT_N/dp_a \approx - 0.14$ K/kbar, $dT_N/dp_c \approx 0.4$ K/kbar, $dT_N/dP \approx  0.13$ K/kbar, so the N\'eel temperature decreases under uniaxial pressure along the $a$-axis and increases (with almost factor of three higher rate) under uniaxial pressure along the $c$-axis.

\begin{figure}[tbp]
\begin{center}
\includegraphics[angle=0,width=120mm]{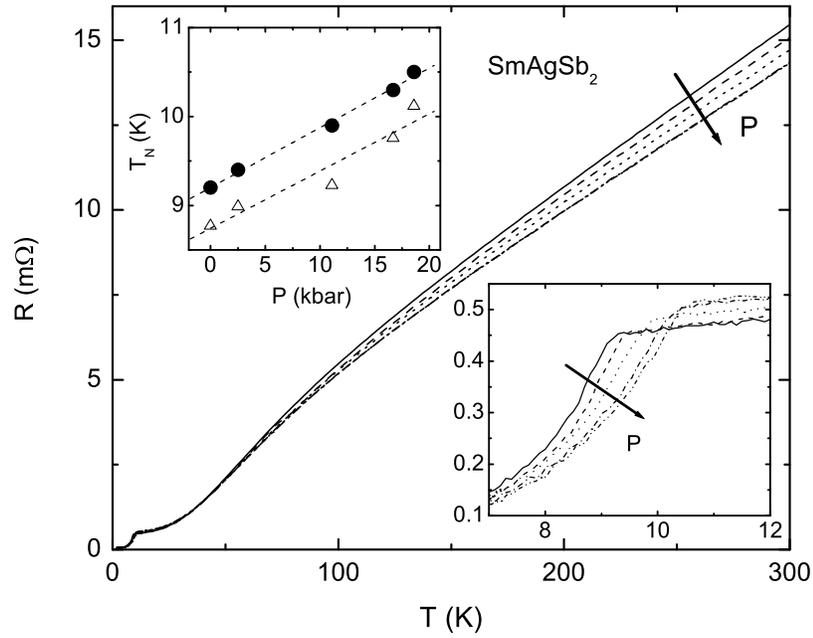}
\end{center}
\caption{(In-plane resistance of SmAgSb$_2$ under hydrostatic pressure (different curves correspond to the pressure values at low temperatures 0, 2.5, 10, 16.7, and 18.6 kbar), the arrow indicates the direction of increasing pressure. Lower right inset: enlarged low temperature part of the main panel. Upper left inset: N\'eel temperature as a function of pressure: circles - from the onset of $R(T)$, triangles from the maximum of $dR/dT$; dotted lines are linear fits to the data.} \label{F6}
\end{figure}

Estimated hydrostatic pressure derivative can be compared with the measured value. Fig. \ref{F6} shows the in-plane resistance of SmAgSb$_2$ as a function of pressure. Room temperature resistivity decreases under pressure with the derivative, $d \ln \rho_{300K}/dP \approx - 4 \cdot 10^{-3}$ kbar$^{-1}$ (room temperature value of pressure were used for this estimate, see e.g. \cite{tho84a} for a discussion of pressure-temperature relation in a piston-cylinder cells). Similar but factor of $\sim 2$ faster decrease of the room temperature resistivity was also observed for LaAgSb$_2$ \cite{tor07a}. At low temperatures, just above the magnetic transition, resistivity of SmAgSb$_2$ increases under pressure and at base temperature, 2 K, it is practically pressure-independent, consistent with the residual resistivity being pressure-independent in our measurements. The N\'eel temperature of SmAgSb$_2$ increases with pressure (Fig. \ref{F6}, upper left inset). Two criteria were used to determine $T_N$: an onset of $R(T)$ and a maximum in $dR/dT$ \cite{fis68a}. The latter criterion gives $T_N$ values consistent with thermodynamic measurements, whereas the former yields slightly higher $T_N$, still both of them give similar pressure derivatives, $d T_N/d P = 0.067 \pm 0.003$ K/kbar and $d T_N/d P = 0.064 \pm 0.01$ K/kbar, for the onset and $dR/dT$ maximum criteria respectively.

The estimate of the $d T_N/d P$ from the Ehrenfest relation is consistent with the direct measurements in its sign but gives a value that is almost a factor of two larger than that measured. However, in terms of absolute amounts this difference is rather small ($< 0.1$ K/kbar) and is probably due to the the accumulation of the error bars from all three measurements.
\\

\begin{figure}[tbp]
\begin{center}
\includegraphics[angle=0,width=90mm]{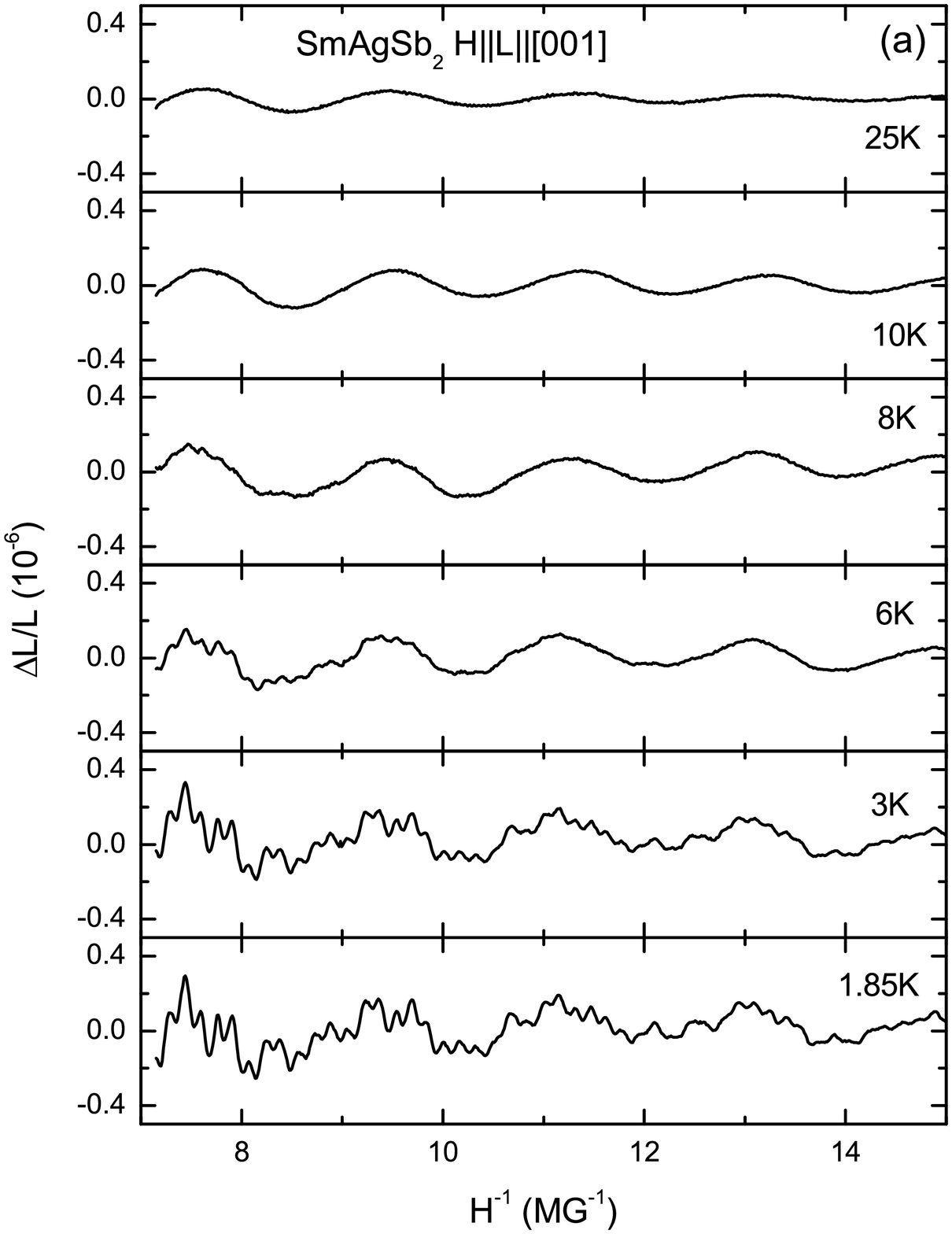}
\includegraphics[angle=0,width=90mm]{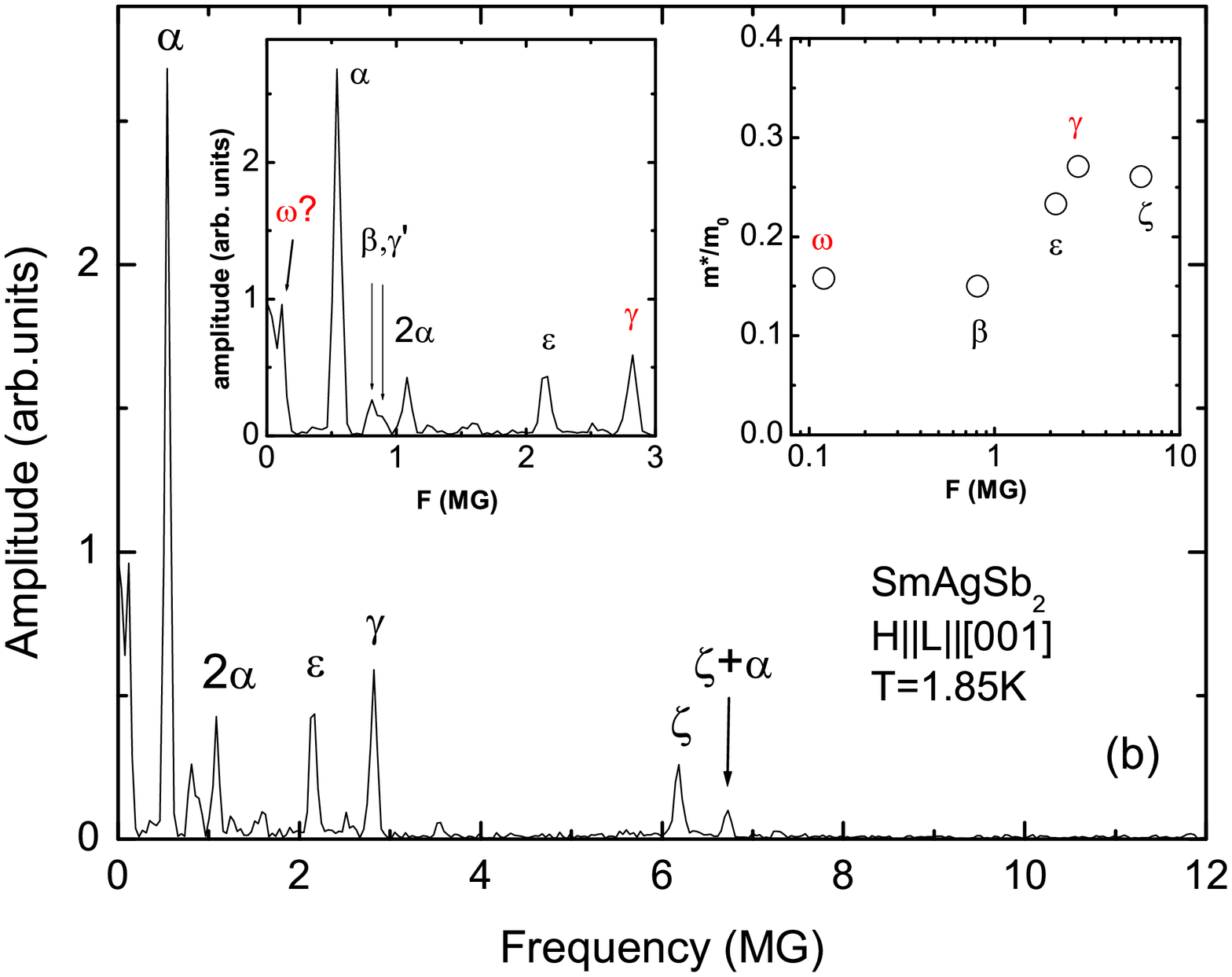}
\end{center}
\caption{(Color online) Oscillations in longitudinal magnetostriction of SmAgSb$_2$ ($H \| L \| [001]$). (a) quantum oscillations of magnetostriction at temperatures up to 25 K plotted as a function of $1/H$ (non-oscillatory background subtracted); (b)fast Fourier transform of the $T = 1.85$ K data in $\Delta L/L_0 (1/H)$ form, see text for labeling of the observed frequencies. Insets to (b): left - enlarged low-frequency part of the graph; right - effective masses of several observed frequencies. New frequencies are marked by red symbols.} \label{F7}
\end{figure}

Quantum oscillations in the longitudinal MS were readily observed for $T \leq 25$ K in SmAgSb$_2$ for $H \| [001]$ (Fig. \ref{F7}(a)). A FFT spectrum of these oscillations at  $T = 1.85$ K is shown in Fig. \ref{F7}(b). The spectra is more complex than that of YAgSb$_2$ (Fig. \ref{F2}) and is in general agreement with the previous works \cite{mye99c,pro07a}. Several details are noteworthy: the dominant frequency in the MS oscillations is $\alpha$, $F_{\alpha} \approx 0.57$ MG, the $\gamma'$ frequency, first reported in Ref. \cite{pro07a}, is seen adjacent to the $\beta$ frequency in the MS measurements as well. A very small, new, frequency, marked here as $\omega$ ($F_{\omega} \approx 0.12$ MG), appears to be present in the spectrum. Band structure calculations usually are not reliable in description of such small FS pockets. Detailed experimental studies are required to unambiguously exclude an artifactitious origin of this frequency. Finally, our data suggest that the frequency (slightly lower than 3 MG) identified in Ref. \cite{mye99c} as a third harmonic of $\beta$-frequency, is actually an independent orbit, marked here as $\gamma$ ($F_{\gamma} \approx 2.82$ MG). The amplitude of this frequency in MS measurements is almost factor of two higher than that of $\beta$, that makes its initial identification as $3\beta$ unlikely \cite{sho84a}. Based on band structure calculations for SmAgSb$_2$ \cite{pro07a} it seems natural to associate this frequency with the $\gamma$ external orbit on the band 1 doughnut-shaped part of the FS. Angular-dependent quantum oscillations measurements are desirable to confirm the assignment of this frequency. Effective masses for several of the frequencies are shown in the right inset to Fig. \ref{F7}(b). The values of $m^*/m_0$ ($m_0$ is a free electron mass) are between 0.1 and 0.3. The effective mass of the $\gamma$-orbit is significantly less than a triple of the $\beta$-orbit effective mass, consistent with our re-identification of the 2.82 MG frequency. Qualitatively similar to the observation in Ref. \cite{mye99c}, the amplitude of the $\alpha$ oscillations as a function of temperature has a break at the temperature corresponding to the $T_N$ in SmAgSb$_2$, whereas the phase of these oscillations does not change at $T_N$ (Fig. \ref{F7}(b)), in contrast to previous findings \cite{mye99c}. Other frequencies were not observed reliably above $T_N$.

\subsection{LaAgSb$_2$}

\begin{figure}[tbp]
\begin{center}
\includegraphics[angle=0,width=120mm]{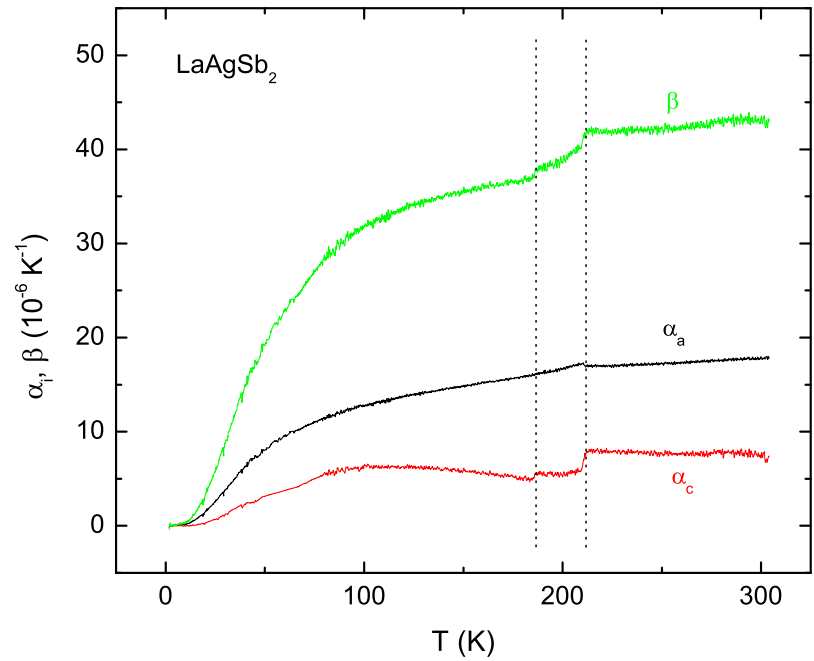}
\end{center}
\caption{(Color online) Anisotropic temperature-dependent linear and volume thermal expansion of LaAgSb$_2$. Dotted vertical lines mark two CDW transitions \cite{mye99a,son03a}.} \label{F8}
\end{figure}

\begin{figure}[tbp]
\begin{center}
\includegraphics[angle=0,width=120mm]{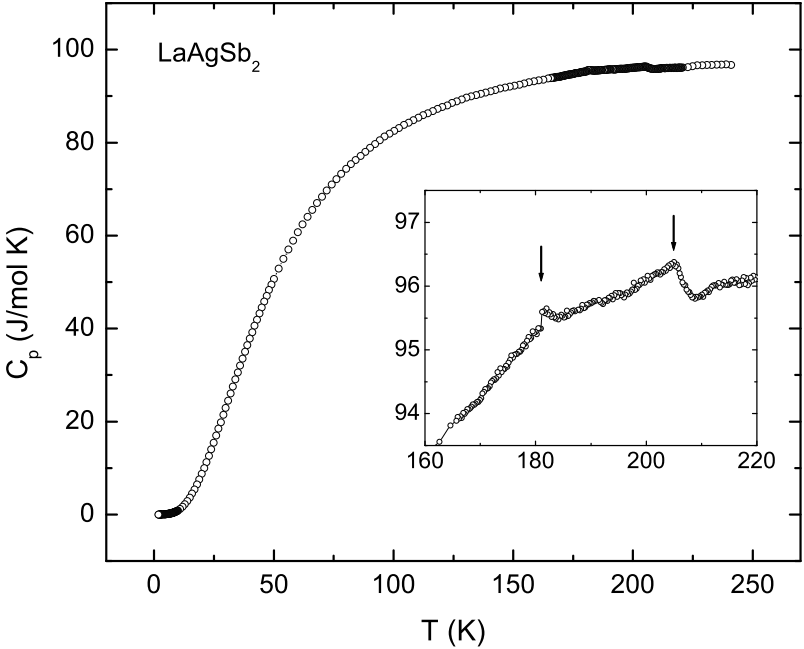}
\end{center}
\caption{Temperature-dependent heat capacity of LaAgSb$_2$. Inset: enlarged part of the region containing CDW transitions. Arrows mark the transitions.} \label{F9}
\end{figure}

The anisotropic, temperature-dependent thermal expansion of LaAgSb$_2$ is shown in Fig. \ref{F8}. Similar to the data for other members of the RAgSb$_2$ family, linear thermal expansion is anisotropic, $\alpha_a > \alpha_c$. Two CDW transitions \cite{mye99a,son03a} are clearly seen in the thermal expansion data: the higher temperature transition manifests itself in both, $\alpha_a(T)$ and $\alpha_c(T)$, whereas the lower temperature transition can be distinguished only in $\alpha_c(T)$; both transitions are seen in the volume thermal expansion, $\beta(T)$. These two CDW transitions are also resolved in heat capacity measurements (Fig. \ref{F9}). We can use the Ehrenfest relation (see above) to estimate the uniaxial and hydrostatic pressure derivatives for these two transitions: for higher temperature CDW: $dT_1/dp_a \approx 1.0$ K/kbar, $dT_1/dp_c \approx -7.2$ K/kbar, $dT_1/dP \approx -5.4$ K/kbar; for lower temperature CDW: $dT_2/dp_a \approx 0$, $dT_2/dp_c \approx dT_2/dP \approx -5.9$ K/kbar. The directly measured hydrostatic pressure derivative for the higher temperature CDW, $dT_1/dP \approx -4.3$ K/kbar \cite{bud06a,tor07a}, is comparable to the one obtained using Ehrenfest relation, there are no direct measurements of $T_2$ under pressure so far. It is noteworthy that both CDW transitions are much more sensitive to the uniaxial pressure along $c$-axis than to that along the $a$-axis and (at least for $T_1$) the signs of the uniaxial pressure derivatives are opposite; additionally, the inferred hydrostatic pressure derivatives are very similar for both CDW transitions, so that a merging of the two transitions is not expected (at least at moderate pressures).

\begin{figure}[tbp]
\begin{center}
\includegraphics[angle=0,width=90mm]{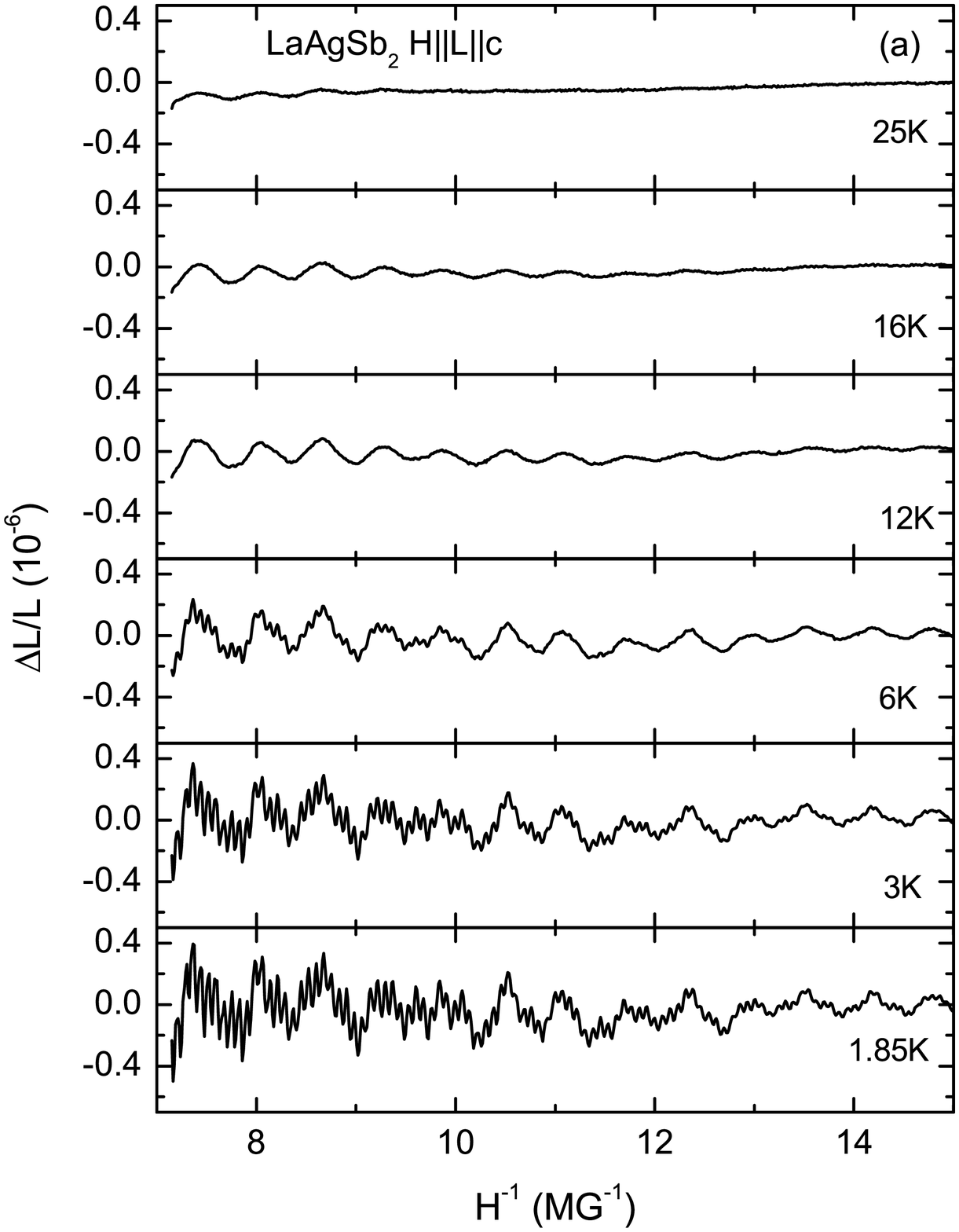}
\includegraphics[angle=0,width=90mm]{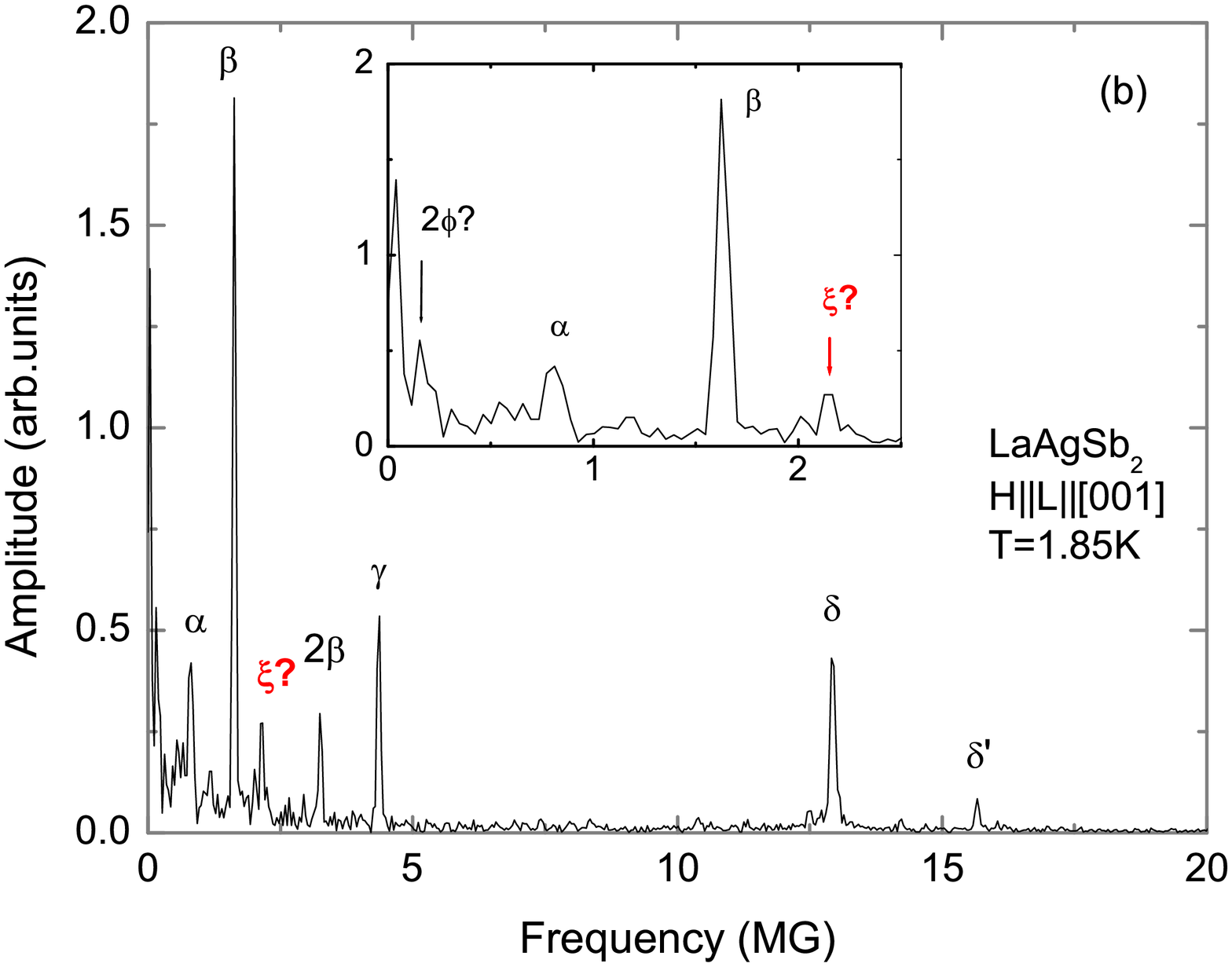}
\end{center}
\caption{(Color online) Oscillations in longitudinal magnetostriction of LaAgSb$_2$ ($H \| L \| [001]$). (a) quantum oscillations of magnetostriction at temperatures up to 25 K plotted as a function of $1/H$; (b)fast Fourier transform of the $T = 1.85$ K data in $\Delta L/L_0 (1/H)$ form, see text for labeling of the observed frequencies. Inset to (b): enlarged low-frequency part of the graph. New frequency is marked by a red symbol.} \label{F10}
\end{figure}

Quantum oscillations in longitudinal MS for $T \leq 25$ K in LaAgSb$_2$ ($H \| [001]$) are shown in Fig. \ref{F10}(a). The extremal orbits observed in magnetostriction are consistent with the ones reported previously \cite{mye99c,ina02a} and are marked on FFT spectrum (Fig. \ref{F10}(b)) using the convention from Ref. \cite{mye99c}. The FFT peak at $\approx 2.14$ MG, labeled $\xi$, possibly corresponds to an extremal orbit that was not previously detected (although a very small peak at a similar frequency can be noticed in a close examination of the FFT spectrum presented in Ref. \cite{mye99c}). The small amplitude of this peak at base temperature does not allow for a determination of its effective mass. Extension of the measurements to lower temperatures is desirable for a clarification of the parameters of this orbit.

\begin{figure}[tbp]
\begin{center}
\includegraphics[angle=0,width=120mm]{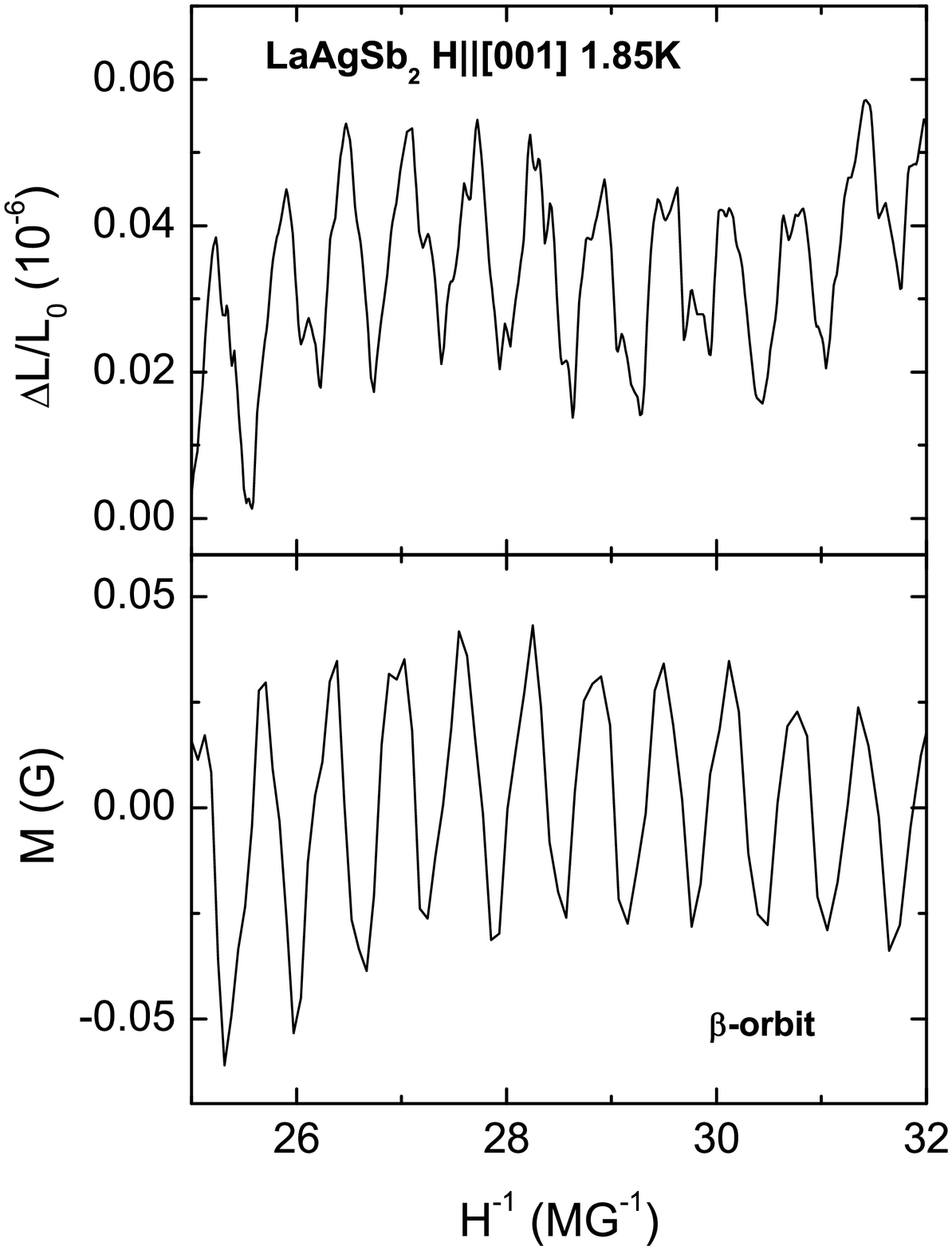}
\end{center}
\caption{Quantum oscillations in LaAgSb$_2$ ($H \| [001], T = 1.85$ K as measured by magnetostriction (upper panel) and magnetization (lower panel).} \label{F11}
\end{figure}

Uniaxial stress dependence of the dominant, $\beta$, frequency is estimated from the comparison of magnetization and magnetostriction measurements (Fig. \ref{F11}) as $\partial \ln S_{\beta} / \partial \sigma_c = - 13 \cdot 10^{-12}$ cm$^2$/dyne, similar to that for $\beta$ orbit in YAgSb$_2$.

\subsection{La$_{0.8}$Ce$_{0.2}$AgSb$_2$ and La$_{0.75}$Nd$_{0.2}$AgSb$_2$}

\begin{figure}[tbp]
\begin{center}
\includegraphics[angle=0,width=90mm]{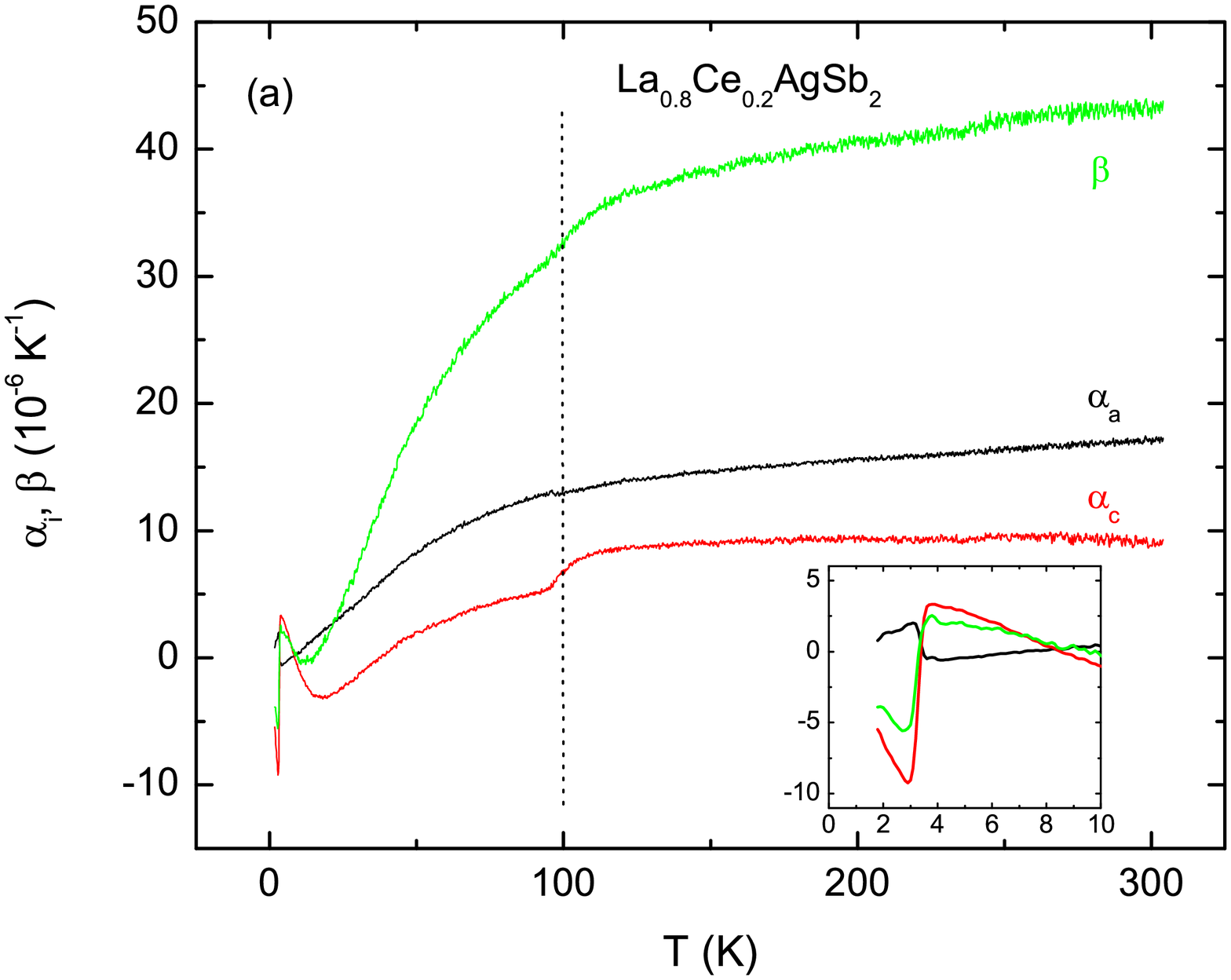}
\includegraphics[angle=0,width=90mm]{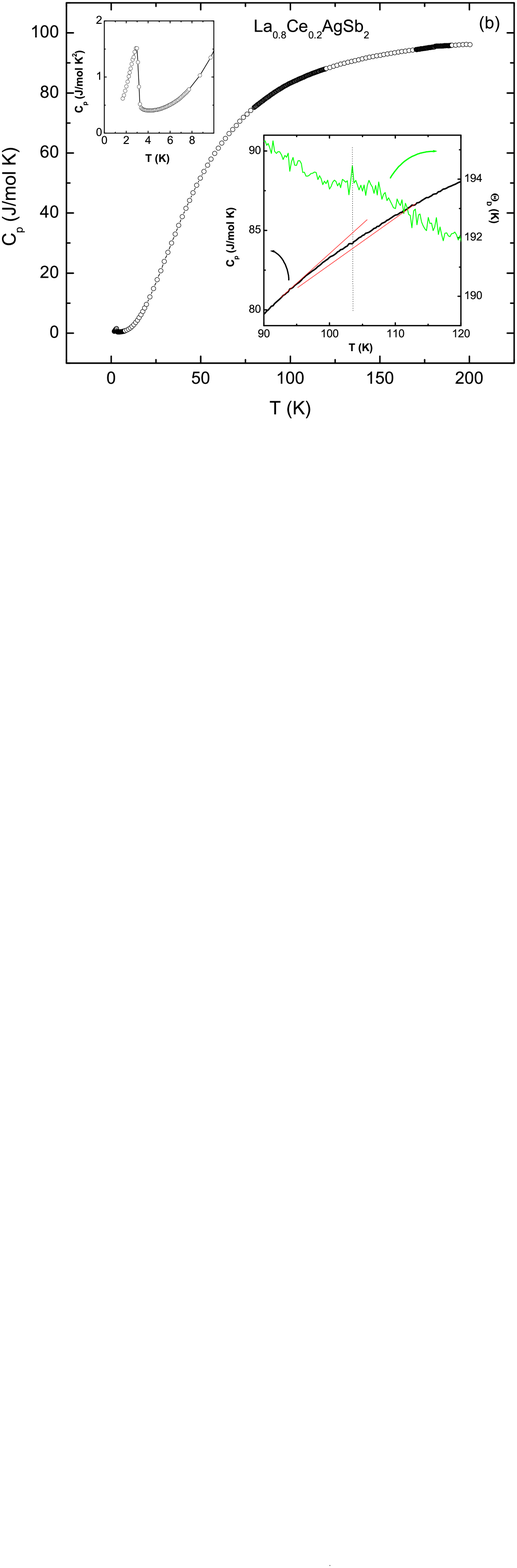}
\end{center}
\caption{(Color online) (a) Anisotropic temperature-dependent linear and volume thermal expansion of La$_{0.8}$Ce$_{0.2}$AgSb$_2$. Dotted vertical line marks the CDW transition. Inset: enlarged low temperature part of the data. (b) Temperature-dependent heat capacity of La$_{0.8}$Ce$_{0.2}$AgSb$_2$. Upper left inset: enlarged low temperature part of the graph. Lower right inset: enlarged part of the $C_p(T)$ graph containing CDW transition and $\Theta_D(T)$ in the same temperature region. Dotted vertical line marks the transition. Red lines are guides for the eye.} \label{F12}
\end{figure}

The anisotropic, temperature-dependent thermal expansion and temperature-dependent heat capacity of La$_{0.8}$Ce$_{0.2}$AgSb$_2$ are shown in Fig. \ref{F12}. Both CDW and ferromagnetic ordering transitions \cite{bud06a,tor07a} are clearly seen in the thermal expansion data with the corresponding features in $\alpha_c(T)$ being significantly larger and of opposite sign in comparison with the features in $\alpha_a(T)$. Expectedly, apart from the features associated with the transitions, the overall temperature dependence of the thermal expansion is an intermediate between the pure LaAgSb$_2$ (see above) and CeAgSb$_2$ \cite{tak03a}. A clear, $\lambda$-shaped, peak in $C_p(T)$ at low temperatures is associated with the ferromagnetic order (Fig. \ref{F12}(b), upper left inset). A feature in the heat capacity corresponding to the CDW transition in this material is practically absent (although it is unambiguous, albeit small, in temperature-dependent equivalent Debye temperature, $\Theta_D(T)$ \cite{bla41a,qua00a}, see Fig. \ref{F12}(b), lower right inset). From the $C_p(T)$ graph (Fig. \ref{F12}(b), lower right inset) we can, very roughly, estimate $\Delta (C_p/T)|_{CDW} \approx 5 \cdot 10^{-3}$ J/mol K$^2$. From the data in Fig. \ref{F12} and the Ehrenfest relation, we can estimate for the ferromagnetic transition: $dT_c/dp_a \approx 0.1$ K/kbar, $dT_c/dp_c \approx -0.46$ K/kbar, $dT_c/dP \approx -0.29$ K/kbar; for CDW: $dT_{CDW}/dp_a \approx 0.7$ K/kbar, $dT_{CDW}/dp_c \approx -6$ K/kbar, $dT_{CDW}/dP \approx -5$ K/kbar. Directly measured \cite{tor07a} hydrostatic pressure derivatives of La$_{0.8}$Ce$_{0.2}$AgSb$_2$ are $dT_c/dP \approx -0.2$ K/kbar (close to the above estimate) and $dT_{CDW}/dP \approx -14$ K/kbar. The absolute value of $dT_{CDW}/dP$ obtained from the Ehrenfest relation is significantly underestimated, probably due to very poor evaluation of $\Delta (C_p/T)|_{CDW}$ (Fig. \ref{F12}(b), lower right inset),  still the existent TE data (keeping in mind that there cannot be ambiguity in the sign of the $\Delta (C_p/T)|_{CDW}$) show that $T_{CDW}$ increases in this material if the uniaxial pressure is applied in the $ab$ plane but decreases for pressure along the $c$-axis and the rate of the change in $T_{CDW}$ is $\sim 9$ times higher for the pressure along $c$, e.g. the response is slightly more anisotropic than for higher temperature CDW in pure LaAgSb$_2$ (see above).
\\

\begin{figure}[tbp]
\begin{center}
\includegraphics[angle=0,width=90mm]{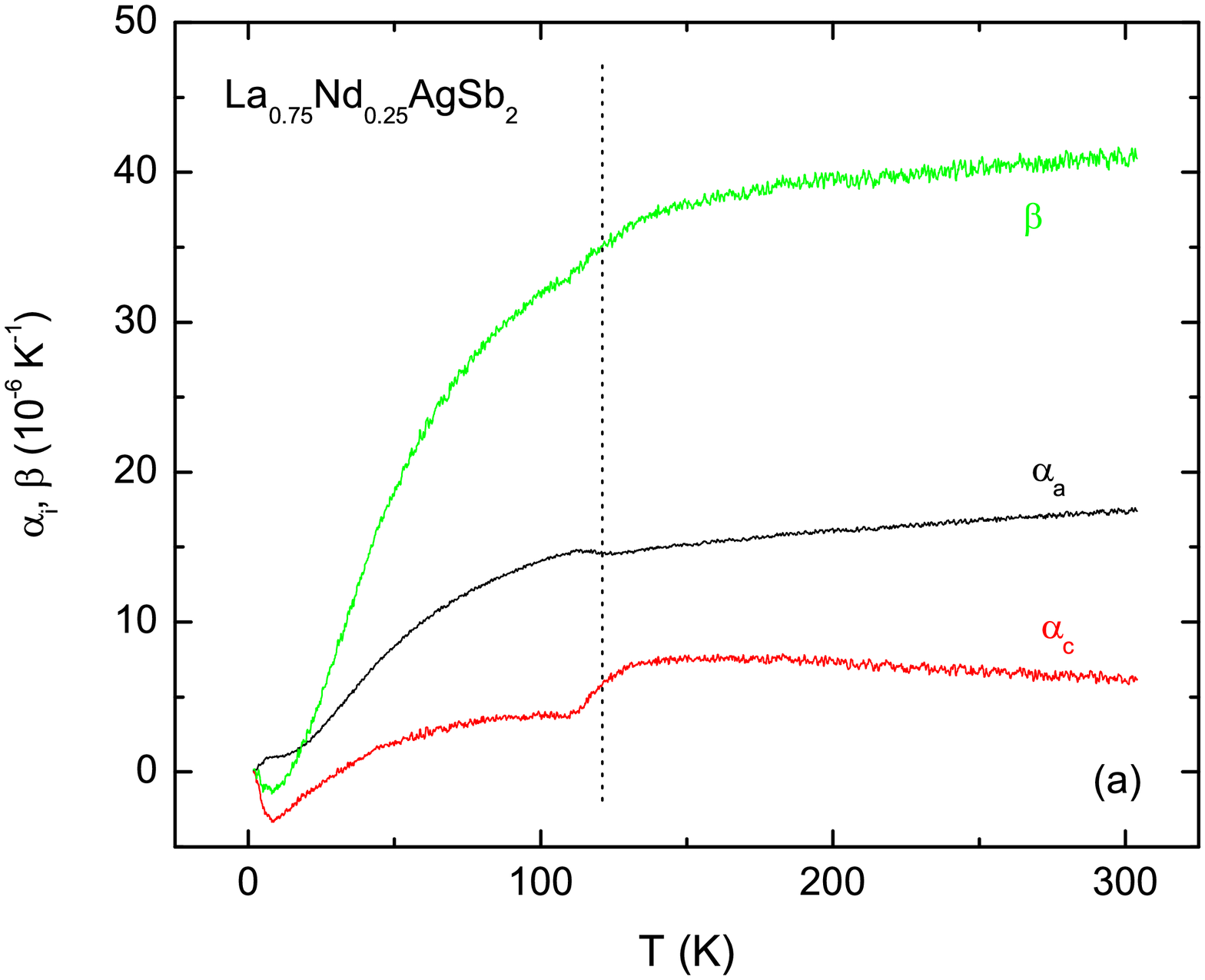}
\includegraphics[angle=0,width=90mm]{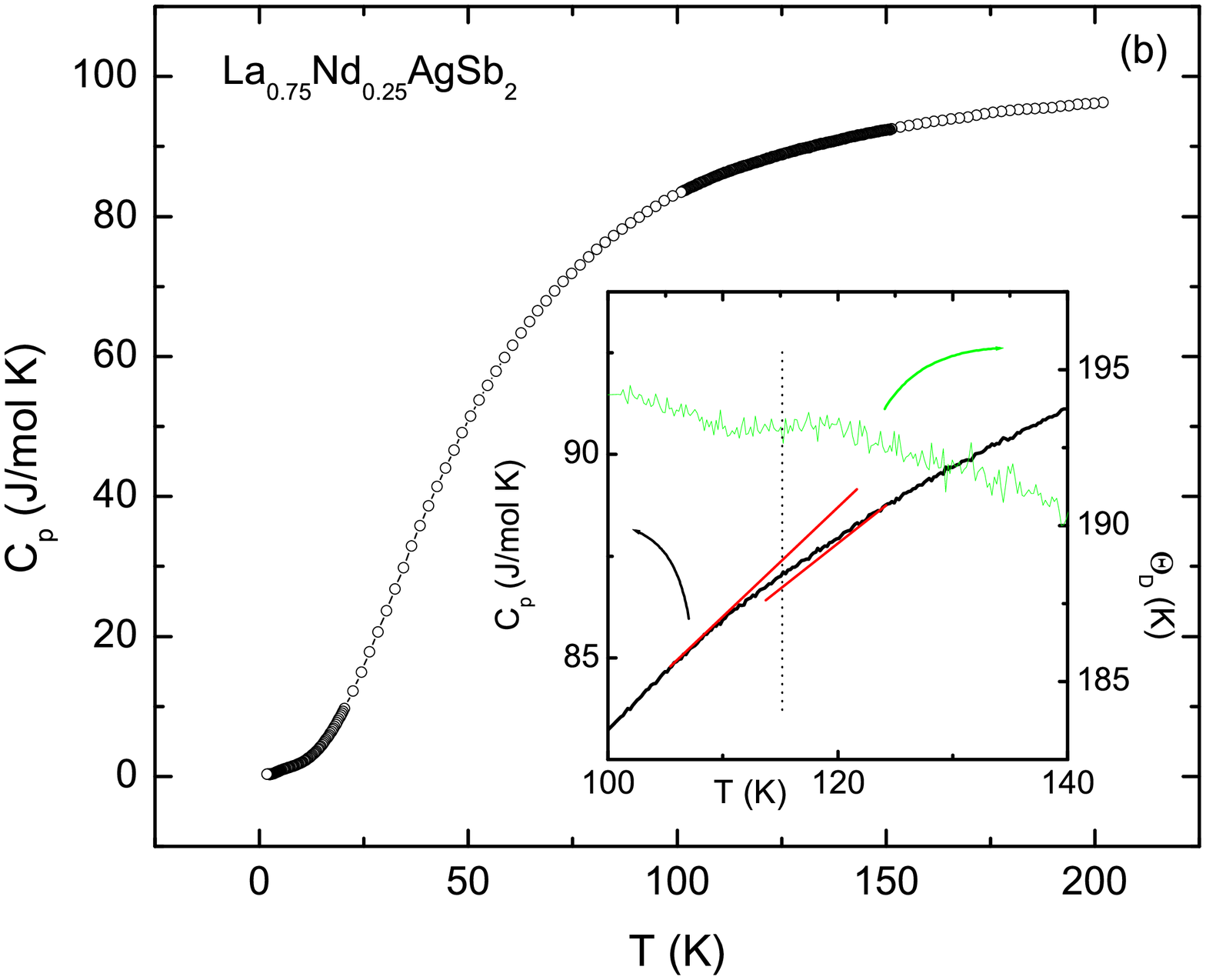}
\end{center}
\caption{(Color online) (a) Anisotropic temperature-dependent linear and volume thermal expansion of La$_{0.75}$Nd$_{0.25}$AgSb$_2$. Dotted vertical line marks the CDW transition. (b) Temperature-dependent heat capacity of La$_{0.8}$Nd$_{0.25}$AgSb$_2$. Inset: enlarged part of the $C_p(T)$ graph containing CDW transition and $\Theta_D(T)$ in the same temperature region. Dotted vertical line marks the transition. Red lines are guides for the eye.} \label{F13}
\end{figure}

A similar set of data for La$_{0.75}$Nd$_{0.2}$AgSb$_2$ is shown in the two panels of Fig. \ref{F13}. This material does not have long range magnetic order (at least above 1.8 K) \cite{bud06a,tor07a}. Broad, low temperature (around 10 K), anomaly in TE and heat capacity is probably associated with the crystalline electric field effects. While CDW transition is clearly seen in TE, similarly to La$_{0.8}$Ce$_{0.2}$AgSb$_2$, $C_p(T)$ data basically have no indication of the CDW transition, even though a weak feature is present in $\Theta_D(T)$ (Fig. \ref{F13}(b), inset). Very roughly we can estimate (an upper limit of) $\Delta (C_p/T)|_{CDW} \approx 5 \cdot 10^{-3}$ J/mol K$^2$. Then $dT_{CDW}/dp_a \approx 2$ K/kbar, $dT_{CDW}/dp_c \approx -9$ K/kbar, $dT_{CDW}/dP \approx -6$ K/kbar. The inferred value of the hydrostatic pressure derivative is very close to the directly measured $dT_{CDW}/dP = -5.7$ K/kbar \cite{tor07a}. The changes in $\alpha_a(T)$, $\alpha_c(T)$ and $\beta(T)$ for are of the same sign but larger that in La$_{0.8}$Ce$_{0.2}$AgSb$_2$ with similar $T_{CDW}$. The anisotropy of the uniaxial pressure response in La$_{0.75}$Nd$_{0.2}$AgSb$_2$ is somewhat smaller than that in  LaAgSb$_2$ and La$_{0.8}$Ce$_{0.2}$AgSb$_2$.

\section{Summary}
The linear thermal expansion in all materials studied in this work is anisotropic, with $\alpha_a > \alpha_c$, apart for the region of magnetic phase transitions, when present. Long range magnetic ordering and CDW transitions present clear anomalies in the thermal expansion data.  Uniaxial pressure derivatives inferred, by using the Ehrenfest relation, suggest that CDW transition temperatures increase when pressure is applied in the $ab$ plane and decrease with pressure along the $c$-axis, the same is true for the ferromagnetic transition in  La$_{0.8}$Ce$_{0.2}$AgSb$_2$, whereas the signs of the uniaxial pressure derivatives are reversed in the case of the antiferromagnetic transition in SmAgSb$_2$. In all cases the effect of the pressure along the $c$-axis is significantly stronger than when the pressure is applied in the $ab$-plane.

de Haas - van Alphen like quantum oscillations in the longitudinal ($H \| L \| [001]$) magnetostriction were observed in the three pure compounds, YAgSb$_2$, SmAgSb$_2$, and LaAgSb$_2$, up to the temperatures as high as 25 K. For the two latter samples new extremal orbits may have been detected.

\ack

Ames Laboratory is operated for the U. S. Department of Energy by Iowa State University under Contract No. DE-AC02-07CH11358. Work at Ames Laboratory was supported by the director for Energy
Research, Office of Basic Energy Sciences. MST was supported by the National Science Foundation under DMR-0306165.
GMS was supported by the National Science Foundation under DMR-0704406. SLB acknowledges McClelland's help in the intermediate stage of this work. All five primary motivations were invoked as driving this work.

\clearpage

\clearpage

\clearpage

\clearpage

\clearpage

\clearpage

\clearpage

\clearpage

\clearpage

\clearpage

\clearpage

\clearpage

\clearpage

\end{document}